\documentclass[twocolumn,preprintnumbers,amsmath,amssymb,aps,prx,a4paper]{revtex4-2}
\usepackage{amsmath}
\usepackage{graphicx}
\usepackage{dcolumn}
\newcolumntype{d}[1]{D{.}{.}{#1}}
\usepackage{bigdelim}
\usepackage{bm}
\usepackage{array}
\usepackage{xcolor}
\usepackage{longtable}
\usepackage{float}
\usepackage{rotating}
\usepackage{multirow}
\usepackage{setspace}

\makeatletter

\newcommand{\Rmnum}[1]{\expandafter\@slowromancap\romannumeral #1@}
\makeatother

\begin{document}

\title{Supra-Binary Ferroelectricity in a Nanowire}
\author{Wentao Xu$^{1}$}
\thanks{These authors contributed equally.}
\author{Lihua Wang$^{2}$}
\thanks{These authors contributed equally.}
\author{Yeongjun Lee$^{3}$}
\author{D. ChangMo Yang$^{2}$}
\author{Amir Hajibabaei$^{2}$}
\author{Cheolmin Park$^{4}$}
\author{Tae-Woo Lee$^{3}$}
\email{twlees@snu.ac.kr}
\author{Kwang S. Kim$^{2}$}
\email{kimks@unist.ac.kr}
\affiliation{$^1$
	Institute of Optoelectronic Thin Film Devices and Technology, Key Laboratory of Optoelectronic Thin Film Device and Technology of Tianjin, Nankai University, Tianjin 300350, P. R. China}
\affiliation{$^2$
	Department of Chemistry, School of Natural Science, Center for Superfunctional Materials, Ulsan National Institute of Science and Technology (UNIST), 50 UNIST-gil, Ulsan 44919, Republic of Korea}
\affiliation{$^3$
	Department of Materials Science and Engineering, Seoul National University (SNU), 1 Gwanak-ro, Seoul 08826, Republic of Korea}
\affiliation{$^4$
	Department of Materials Science and Engineering, Yonsei University, 50 Yonsei‐ro, Seoul 03722, Republic of Korea}
\date{\today}

\begin{abstract} 
We report the prediction and observation of supra-binary ferroelectricity in a ferroelectric nanowire (FNW) covered with a semi-cylindrical gate that provides an anisotropic electric field in the FNW. There are gate-voltage-driven transitions between four polarization phases in FNW's cross section, dubbed axial-up, axial-down, radial-in and radial-out. They are determined by the interplay between the topological depolarization energy and the free energy induced by an anisotropic external electric field, in clear distinction from the conventional film-based binary ferroelectricity. When the FNW is mounted on a biased graphene nanoribbon (GNR), these transitions induce exotic current-voltage hysteresis in the FNW-GNR transistor. Our discovery suggests new operating mechanisms of ferroelectric devices. In particular, it enables intrinsic multi-bit information manipulation in parallel to the binary manipulation employed in data storage devices. 
\end{abstract}

\pacs{74.20.De, 77.80.-e, 77.80.Bh, 77.84.-s, 85.50.Gk}
\maketitle

A ferroelectric material~\cite{Wang2003,Horiuchi2008,Zhang1998} exhibits persistent polarization when exposed to a weak external electric field. As the field passes some threshold values, the polarization undergoes a first-order transition between different phases~\cite{Ginzburg1945} and the reported transitions have been mostly binary. 
Memory devices built from these materials are being proposed as new approaches to non-volatile random-access memory, in which arrays of ferroelectric cells store binary information which can be altered in response to an external electric field.
Any improvement of the storage capacity thereupon would involve increasing the number of units. Given the quantum limit which foretells the impending curtailment of such growth, exploring the possibility of storing multiple bits per unit is considered as an alternative. Recent non-material innovations have demonstrated multi-bit data storage units with the conventional binary ferroelectricity, by combining translational ferroelectric domain strips~\cite{Tripathi2011} or many ferroelectric nanoparticles~\cite{Sohn2010}.  Very recently, ternary metal oxide semiconductor technology based on tunneling appeared~\cite{Jeong2019}.  These approaches improve the ratio of read-and-write heads to storage units while limited by the same areal scaling of the storage capacity. To improve the areal scaling requires a theoretical re-examination of the origin of the ferroelectric phase transition, on which numerous previous studies exist. First-principles calculations~\cite{Zhong1995} have been used to explain the ferroelectricity in well-defined lattice structures such as ABO$_3$ materials. Exploration of ferroelectric properties of organic polymers has been made by evoking microscopic processes such as charge transfer~\cite{Tayi2012}. 

Meanwhile, the phenomenological and macroscopic Landau-Ginzburg theory (LGT) is used for general ferroelectric materials~\cite{Kretschmer1979}. And, the discovery of ferroelectric property by LGT has never ceased. For instance, the Landau double-well energy landscape in a binary ferroelectric layer was recently unveiled~\cite{Hoffmann2019}. Of interest in this work are composition-oblivious supra-binary ferroelectric properties, for which LGT is extended. We predicted and observed the supra-binary ferroelectricity in a ferroelectric nanowire (FNW) covered with a semi-cylindrical gate that provides an anisotropic electric field in the FNW. There are gate-voltage-driven transitions between four polarization phases in FNW’s cross section, dubbed axial-up, axial-down, radial-in and radial-out.

In LGT, a dipole density field $\vec{P}\left(\vec{r}\right)$ is used to describe the polarization in a ferroelectric material. The Gibbs free energy $\mathcal{F}$ is built upon this field. The system must settle into one of minima of $\mathcal{F}$, thus, yielding a specific polarization and breaking the symmetry spontaneously. The minima and the energy barrier between them evolve with the external electric field. A phase transition of polarization happens when the system overcomes the energy barrier to choose another minimum of $\mathcal{F}$. We show that the Gibbs free energy term due to depolarization may switch on or off depending on the topology of $\vec{P}\left(\vec{r}\right)$ and its domain. This switch, along with the interaction between dipole density field and external electric field, may induce new phase transitions that provide candidates for the multibit per unit, in a FNW.

\begin{figure}
	\begin{center}
		\includegraphics[width=1.\columnwidth]{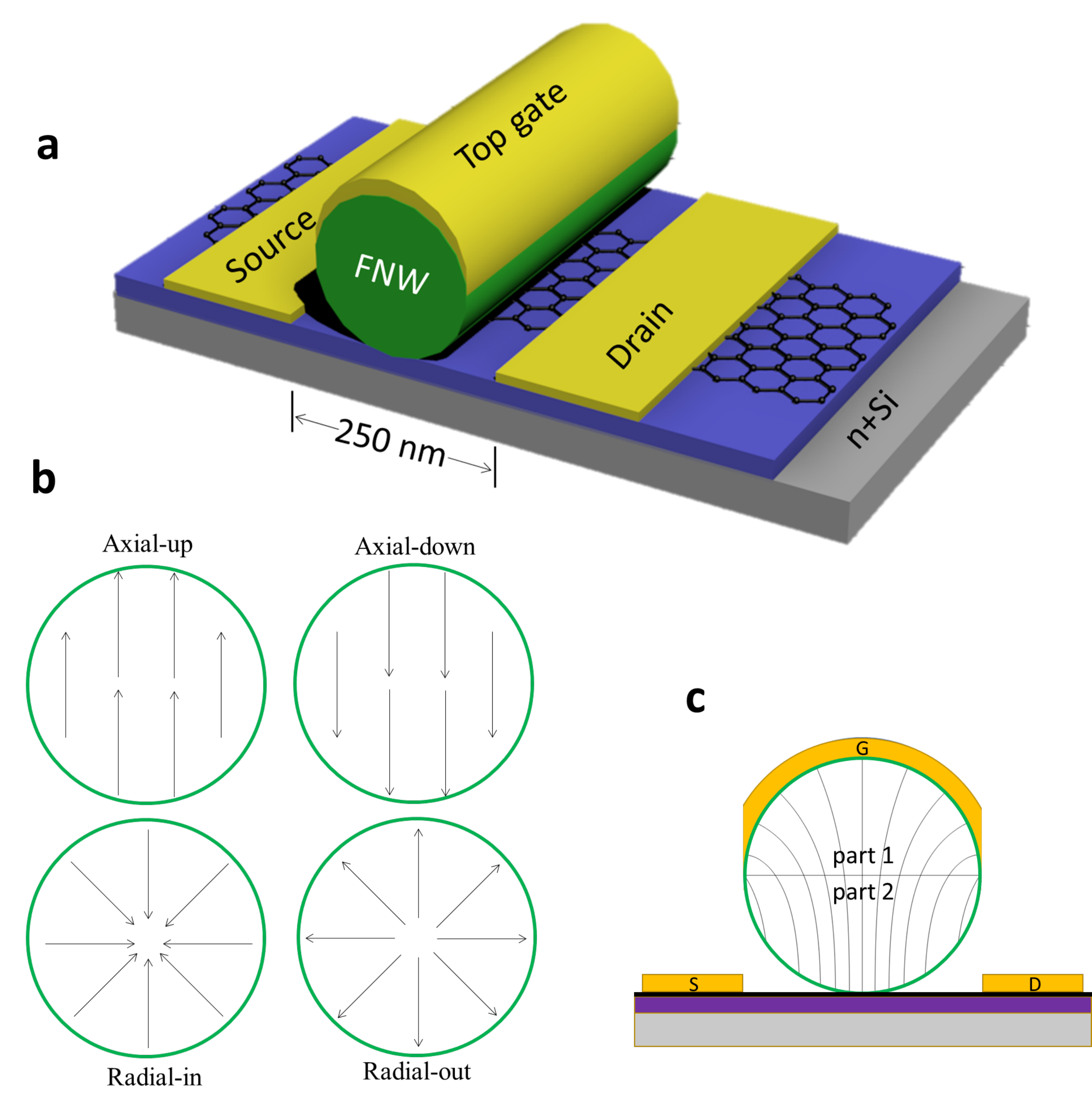}\\
		\caption{\label{fig:setup} Schematic of the one-step self-aligned GNR FET using FNWs with semicircular top-gate. a, Pre-aligned GNRs were patterned from CVD-grown graphene by using organic NW lithography. PVDF$_{0.7}$-TrFE$_{0.3}$ FNW is used as the ferroelectric gate insulator on which continuous gate was formed. PVK NW is used as separator. A 40-nm Au layer was self-aligned in one step during evaporation into source (S), drain (D) and gate (G) electrodes. b, The four polarization phases in FNW’s cross section. c, Electrical field configuration in the cross-section of the FNW, simulated by the finite element method.}
	\end{center}
\end{figure}

To test this possibility, we used an elecrohydrodynamic nanowire printer to draw long straight FNWs on a large area with individual control over positioning and alignment, perpendicular to underlying GNR. In one step, semi-cylindrical metallic electrodes on top of FNW and two contact pads under the FNW's rims were self-aligned. A field-effect transistor (FET) was then built, combining a top-gated FNW and a pre-etched biased GNR underneath it. We chose the organic copolymer P(VDF-TrFE) as the printed material~\cite{Zhang1998}. Its nonlinear nature and easy processability provide great potential for ferroelectricity~\cite{Kawai1969,Bergman1971}. Devices were fabricated using all-nanowire-templated lithography and self-alignment approaches (Fig. \ref{fig:setup}). To avoid additional contaminants that complicate electrical characteristics, the process did not use traditional lithography. A graphene sheet was synthesized by chemical-vapor deposition on Cu foil, and then transferred to a highly-doped silicon wafer coated with a $300$-nm SiO$_2$ layer~\cite{Bae2010}. An array of parallel polyvinyl carbazole (PVK) NWs with a pitch of $50$ $\it{\mu}$m was printed on the graphene sheet to provide protective masks~\cite{Min2013,Xu2014}. Elecrohydrodynamic nanowire printing produces NWs with nearly perfect circular cross-section~\cite{Min2013}. O$_2$ plasma was used to etch away the unprotected regions to leave GNRs beneath the narrow contact between cylindrical NWs and graphene. PVK masks were removed by brief sonication in chloroform to leave GNRs with individual width $\sim$30 nm. A second series of PVK NWs was then printed in parallel but with staggered GNRs, which was later served as separators of metal pads. Constant-pitched PVDF$_{0.7}$-TrFE$_{0.3}$ FNWs with diameter $\sim$250 nm Supplemental Material Figs. S1, S2) in $\beta$-phase (Supplemental Material Fig. S3) were printed on the same plane but vertically to the longitudinal directions of GNRs. A thin metal layer (Ti/Au 3 nm/35 nm) was then deposited by thermal evaporation, and the layer was directly separated into several regions (Supplemental Material Figs. S4, S5), in which four pads are in direct contact with the substrate, with two metallic lines on PVK NWs, and with one metallic line on PVDF$_{0.7}$-TrFE$_{0.3}$ FNW. Thus, the FNW is covered with a semicircular cross-sectioned electrode. Consequently, the source, drain, and gate electrodes were formed in one step. Two of the pads in contact with two ends of GNRs functioned as source and drain electrodes; the channel length was approximately the diameter of the FNW. 

Most existing studies used P(VDF-TrFE) as a thin film; attempts to explore its ferroelectric switching mechanisms and electronic applications in nanowires (NWs) are rare, due to the lack of both theoretical foundations and appropriate fabrication techniques. Our work fills this gap. In result, we observed exotic current-voltage ($\it{I}$-$\it{V}$) responses on GNR, which are attributed to novel transitions among four phases by both radial and axial alignment of the dipole density field in the FNW. Because these transitions are distinct from the binary phase transitions in traditional thin-film ferroelectrics, the mechanisms of the phase transition in the FNW and GNR's $\it{I}$-$\it{V}$ response to the transition must be explored theoretically. We propose a theoretical model that explains well the observed phenomenon.

We will assume translational symmetry along the longitudinal direction of the FNW and confine our discussion to the cross-section of the FNW. The metallic gate line provides the FNW with an anisotropic electric field (Supplemental Material Methods: determination of the external electric field) which is radial towards the upper half circle (part 1), but axial in the lower half circular region (part 2). Numerical results first show that the dipole field will switch between radial-in and radial-out in the FNW when the radial external field $\vec{E}_{\it{ext}}$ varies~\cite{Hong2008,Note1} and then predict spontaneous radial polarization in the FNW~\cite{Hong2010}. The latter was confirmed in experiment~\cite{Ahluwalia2013}. In fact, we point out that a spontaneous radial polarization in the FNW can be explained within LGT. The depolarization field $\vec{E}_{\it{D}}$ is switched on or off in the FNW depending on the topology of dipole density field. We recall that $\vec{E}_{\it{D}}$ is induced by the charge on the inner surface resulting from the dipole-surface projection $\hat{S}\cdot \vec{P}\left(\vec{r}_s\right)$ where $\hat{S}$ is the unit normal vector at a surface point $\vec{r}_s$ (Supplemental Material Methods: phase transition among four polarization phases in the FNW). Because radial polarization in a circular domain is the uncommon case in which $\vec{E}_{\it{D}}$ is switched off without any surface charge compensation, this switch is topological. In other topologies the dipole field interacts with both $\vec{E}_{\it{ext}}$ and $\vec{E}_{\it{D}}$. While $\vec{E}_{\it{D}}$ is rigidly controlled by the magnitude $P_0$ of spontaneous polarization, $\vec{E}_{\it{ext}}$ varies widely with gate voltage $V_G$. Therefore, the two interactions give competing contributions to the Gibbs free energy to enable novel phase transitions in the FNW (Supplemental Material Methods: phase transition among four polarization phases in the FNW, Supplemental Material Animation 1).
  
\begin{figure}
	\begin{center}
		\includegraphics[width=1.\columnwidth]{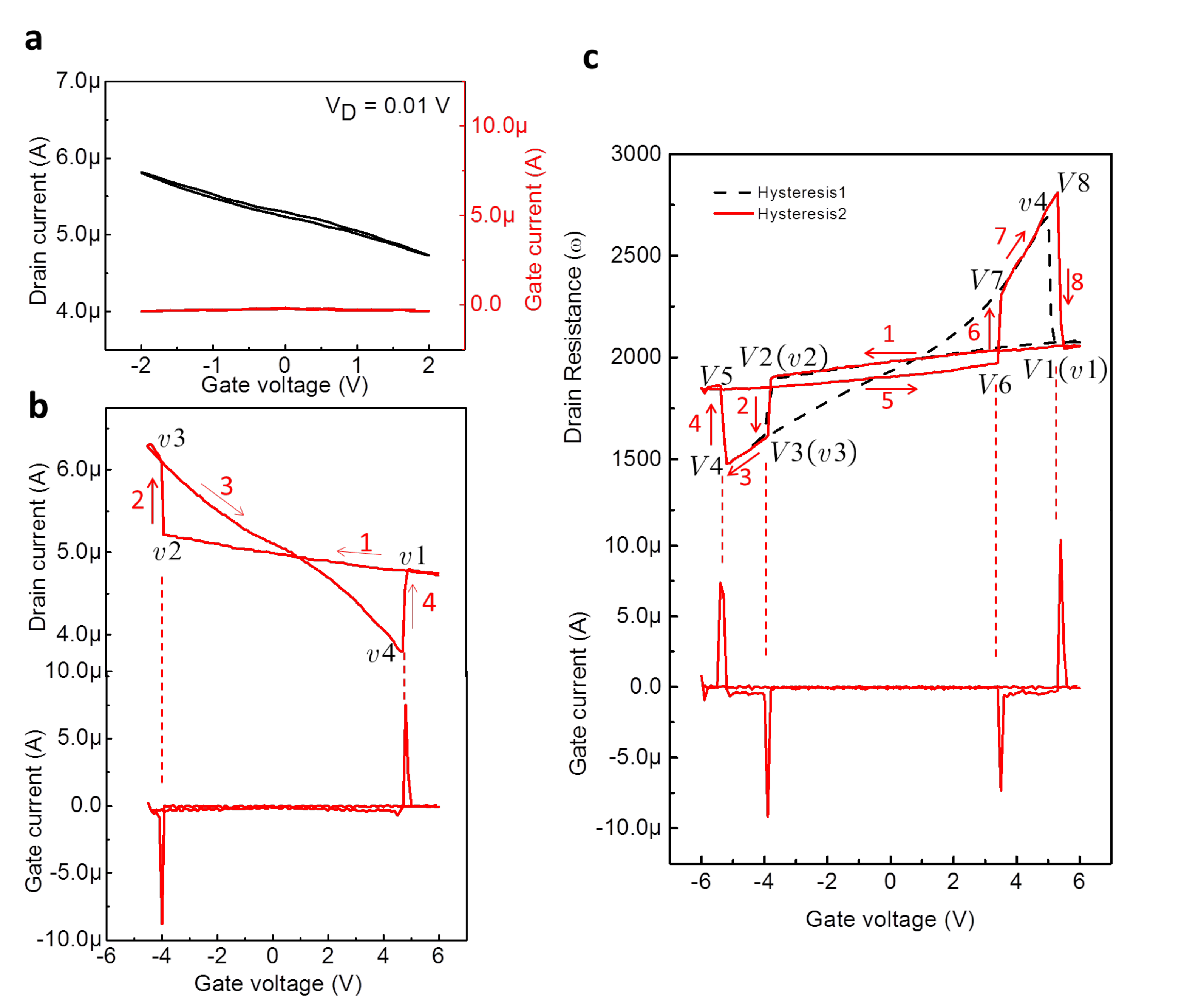}\\
		\caption{\label{fig:hysteresis} Characteristics of electronic device in top-gated architectures and electrical responses to the gate voltage sweep. a, No phase transition exists for the low voltage sweep range between 2V and -2 V. b, Drain current (upper) and gate current (lower) for the moderate gate voltage swept (Type II) between +6 V and -5 V. Two phase transitions occur, one in each of stages 2 and 4. c, Drain resistance (upper solid hysteresis2 in red) and gate current (lower) for the large gate voltage swept (Type III) between +6 V and -6 V. Four phase transitions occur, one in each of stages 2, 4, 6 and 8. Dashed hysteresis 1 in black refers to that of type II sweep for reference. See also Supplemental Material Animation 2.}
	\end{center}
\end{figure}

The FNW shows four phases of polarization in the cross section of the FNW during sweeps of $V_G$: radial-in, radial-out, axial-up, and axial-down. Depending on the sweeping range and manner, it undergoes type I: no phase transition (-2 V to 2 V, Fig. \ref{fig:hysteresis}a); type II: transition between two phases (-5V to 6V, Fig. \ref{fig:hysteresis}b); type III: transition among all phases (-6V to 6V, Fig. \ref{fig:hysteresis}c).
\begin{figure*}
	\begin{center}
		\includegraphics[width=0.65\textwidth]{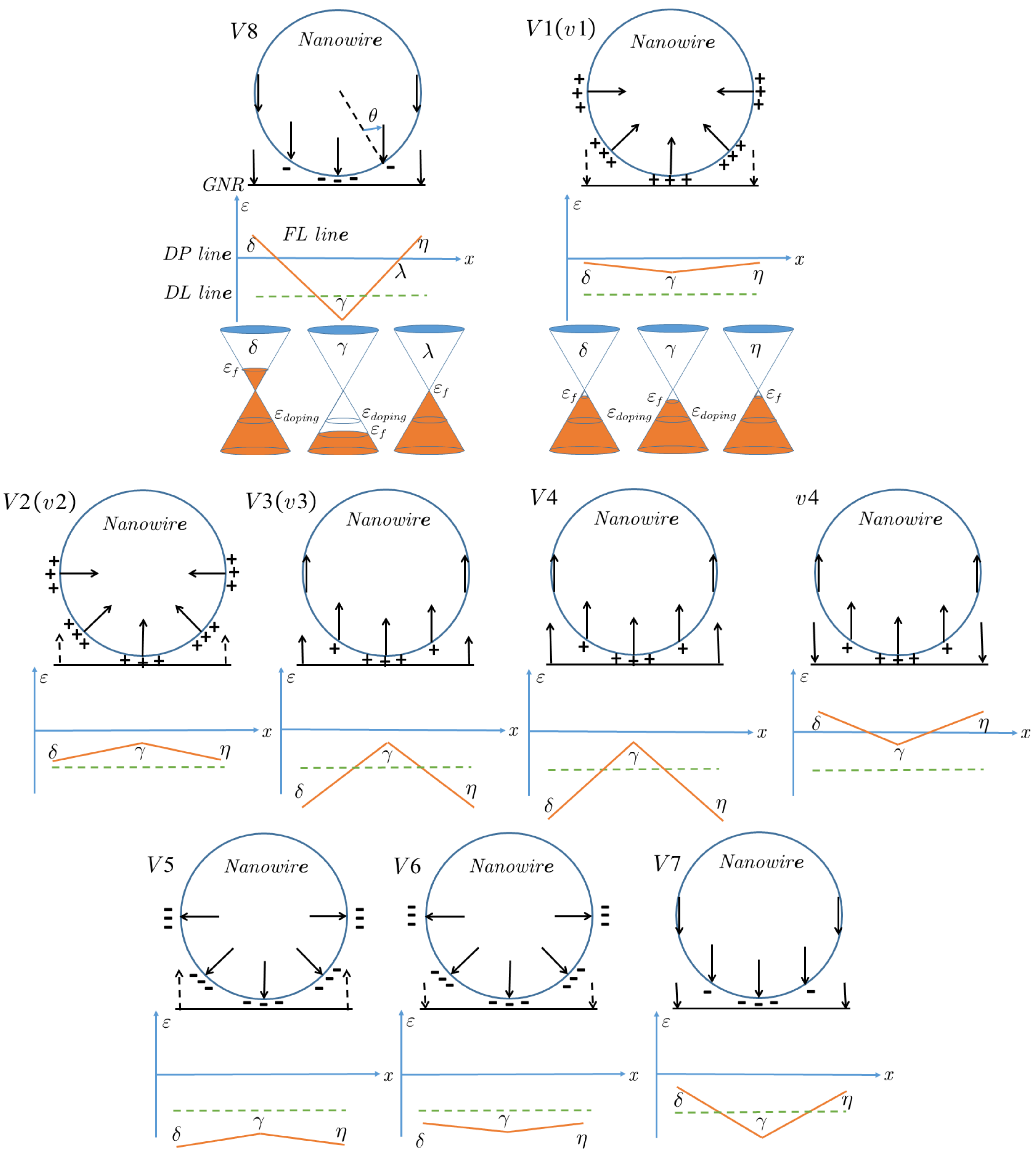}\\
		\caption{\label{fig:explanation} Band diagrams, including the Fermi level (FL), Dirac point (DP), and doping level (DL) lines, explains GNR resistance in various polarization phases of the FNW.  Solid arrow inside of FNW: dipole moment. Solid arrow on GNR: non-screened gate field. Dashed arrow on GNR: almost-screened gate field. +/- symbol: the surface charge. The coordinate frame: the x-axis is the DP line while the y-axis is the energy level. Green dashed line: the DL line below the DP line for p-doping. The FL line in Orange is determined by distinct combinations of the gate field and polarization phases of the FNW. The enclosed area between FL and DP lines determines the conductivity on GNR, giving vortices of the GNR’s $\it{I}$-$\it{V}$ hysteresis shown in Fig.\ref{fig:hysteresis}: v1-v4 for moderate   sweep (type II) while V1-V8 for the large sweep (type III). In an axial phase (e.g., V8), the gate field below the rim of FNW’s diameter is non-screened. Dispersions show that the FLs at the two end points   and   are in the conducting band while that at the middle point   is in valence band. FL and DP lines must cross at point  . But in a radial phase (e.g., V1(v1)) the surface charge is uniform, screening the gate field onto GNR. In result, the FL line is almost flat with the similar dispersions at the ends and in the middle of GNR. See also Supplemental Material Animation 2..}
	\end{center}
\end{figure*} 
In contrast, the GNR's $\it{I}$-$\it{V}$ response is affected delicately by the phase transitions in the FNW. First, the electric field on the GNR consists of (a) surface field that is created by the domain’s surface charge $-\alpha \hat{S}\cdot \vec{P}\left(\vec{r}_s\right)$ which resides on the domain’s outer surface and is simplified as a partially compensating charge to the depolarization field~\cite{Kretschmer1979,Chen1999} with a factor $0<\alpha<1$ and (b) gate field that is generated by the electrode. The surface field is determined by the polarization phase in the FNW, and the gate field is also affected indirectly by a switching of surface-charge screening (Supplemental Material Fig. S7). Band diagrams (Fig. \ref{fig:explanation}) illustrate the influence of the electric field on the GNR under the FNW. The Fermi-level (FL) line connects FL in each segment of the GNR approaching the continuum limit. The FL line is drawn in a simplified manner by connecting the FLs of its middle point $\gamma$ and end points $\delta$ and $\eta$. Similarly, the Dirac-Point (DP) line connects every DP located at zero for each segment of the GNR approaching the continuum limit. Positively shifted Dirac voltage suggests that the GNR is subject to a strong p-doping which is expected when graphene is exposed to moisture and oxygen in air (Supplemental Material Fig. S6). The doping-level (DL that is FL in the presence of chemical doping) line, which connects DLs for each segment of the GNR approaching the continuum limit, becomes a horizontal line below the DP line when doping is p-type; the DL line is the reference line from which the FL line is affected by the electric field. When the dipole field is radial, the surface charge is uniform (Fig. \ref{fig:explanation}: V2). The gate field is almost entirely screened (Supplemental Material Fig. S7: V2). Thus, the FL line is flat except that points $\delta$ and $\eta$ are slightly lifted or lowered according to the polarity of $V_G$. In turn, the FL bends slightly by an amount related to the magnitude of $V_G$. However, when the dipole field is axial, the surface charge density decays quickly with increasing distance from point $\gamma$ (Fig. \ref{fig:explanation}: V3). Thus, although the point $\gamma$ of a FL line is still determined by the surface charge, points $\delta$ and $\eta$ of a FL line are much affected by the unscreened gate field below the rim of FNW's diameter (Supplemental Material Fig. S7: V3). Hence the FL line is steeply bent. 

\begin{figure}[h!]
	\begin{center}
		\includegraphics[width=1.\columnwidth]{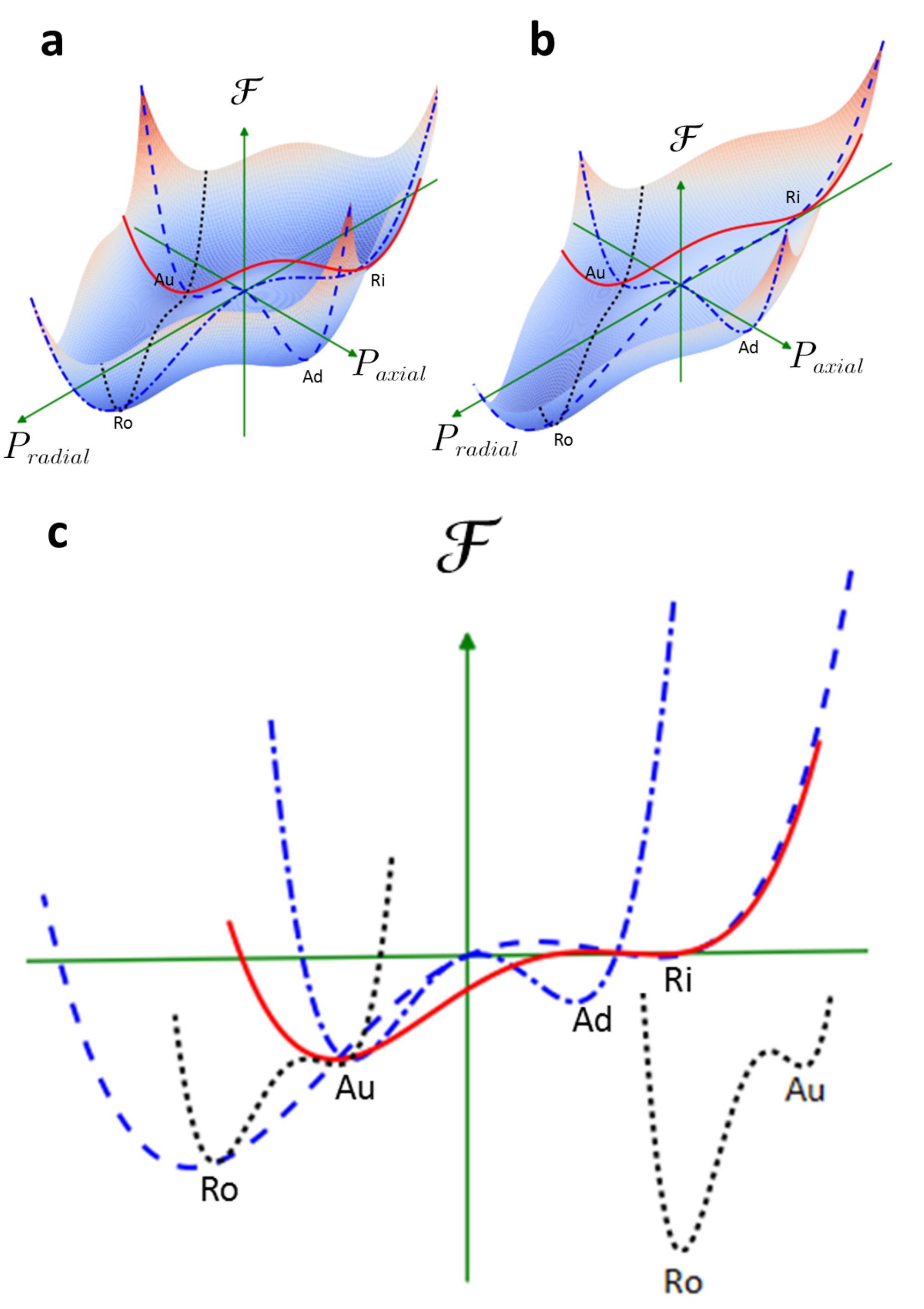}\\
		\caption{\label{fig:energy-surface} Schematic of Gibbs free energy surface defined in an abstract polar coordinate frame. The radial coordinate refers to the characteristic polarization magnitude while the orientation of the dipole field varies with the angular coordinate. The four minima correspond to the radial-in (Ri), radial-out (Ro), axial-up (Au), and axial-down (Ad) polarization phases in the FNW. The dashed energy curve connects Au and Ad; the dot-dashed curve connects Ro and Ri. Evolving along these two curves determines the traditional binary phase transition. The solid curve (Red) connects Ri and Au while the dotted curve connects Au and Ro. Evolving along these curves that connect phases of different orientations determines the supra-binary phase transitions. a, Undisturbed energy surface. Ro and Ri are equally deeper (in colder color) and separated more widely. The state cannot overcome any energy barrier. The spontaneous polarization of the FNW is in one of the two radial phases. b, The energy surface is tilted when VG = -4 V (V2-V3 in Fig. \ref{fig:hysteresis}c), by the bilinear interaction between the electric and dipole fields which differs for radial and axial polarizations in an anisotropic electric field. c, The side view of the four energy curves of (b). The energy barrier between Ri and Au dissolves so that a transition from Ri to Au suddenly occurs. The inset, a clearer view of the dotted curve, shows the energy barrier between Au and Ro prevents further transition from Au and will dissolve to enable another transition from Au to Ro until VG = -5.5 V (V4-V5 in Fig. \ref{fig:hysteresis}c). Supplemental Material Animation 1 further illustrates all transitions between two phases and between four phases for moderate electric filed sweep (type II) and large sweep (type III) respectively.}
	\end{center}
\end{figure} 
The combination of two directions of gate field and four orientations of dipole field yield eight distinct distributions of electric field longitudinally along GNR; together with the chemical doping, such field distributions determine eight distinct FL lines of GNR. In contrast, the electron-hole symmetry in the conductivity gives a unimodal mapping from FL to conductivity, and thereby yields a hysteresis loop, that has up to eight vortices and has switching chirality, in GNR's $\it{I}$-$\it{V}$ response. The loop has a butterfly shape with four vortices (v1-v4 in Fig. \ref{fig:hysteresis}b,c) when involving two phases~\cite{Drincic2011}, but a more complicated ‘fan-blade’ shape with eight vortices (V1-V8 in Fig. \ref{fig:hysteresis}c) when involving four phases. (A detailed walkthrough of both the butterfly and ‘fan-blade’ hysteresis loops is presented in Fig. \ref{fig:energy-surface}, Supplemental Material Methods: hysteresis loop walkthrough and Supplemental Material Animation 2).

Qualitative features in the $\it{I}$-$\it{V}$ hysteresis arise directly from two events. First, whenever the FL and DP lines cross, the resistance increases quickly, as if a blockage occurs in the corresponding GNR segment to dominate the total resistance but not to completely block the circuit due to the intrinsic conductivity at a finite temperature~\cite{Neto2009,Xu2015}. Second, the doping type determines the overall orientation of the hysteresis loop because it will determine whether the lifting/lowering of the FL will increase/decrease the conductivity. 

In conclusion, we observed and explained new transitions among four phases of the polarization in a semicircular top-gated FNW which forms a unique FNW-GNR-FET together with an underlying biased GNR. The phase transitions are explained within LGT. The consequent exotic hysteresis of GNR’s $\it{I}$-$\it{V}$ response is explained graphically by the interplay of FL, DP and DL lines. These new ferroelectric phase transitions may stimulate a new field of ferroelectric research. For instance, the four polarization phases in FNW provide intrinsic multi-bit in a non-volatile memory and the switching of the hysteresis chirality in a FNW-GNR-FET has potential in algebraic operation of multi-digit information~\cite{Kim2017}. The easy all-nanowire-based and self-aligned fabrication approach used to make FET is an important step toward low-cost commercialization of highly sophisticated nano-electronics. Given the boost in the storage capacity by the intrinsic multi-bit information storage with supra-binary ferroelectricity, our findings represent a major step forward in device miniaturization which includes a reduction of read-and-write heads as well as the physical size of the storage media.

\clearpage
\begin{acknowledgments}
This research was supported by the Pioneer Research Center Program through the National Research Foundation of Korea funded by the Ministry of Science, ICT and Future Planning (2012-0009460), the Center for Advanced Soft Electronics funded by the Ministry of Science, ICT and Future Planning as Global Frontier Project (2014M3A6A5060947), and National Research Foundation (National Honor Scientist Program: 2010-0020414) of Korea. 

W.X.~designed, organized and conducted most of the experiments, including fabrication of devices and analysis of data. L.W.~constructed the physical model, predicted and explained the ferroelectric phase transitions and the hysteresis in GNR I-V characteristics. A.H.~contributed to the Gibbs free energy analysis. D.C.Y.~contributed to the electric field simulation. Y.L.~contributed to the fabrication of fluorinated NW. C.P.~commented the paper. W.X.~initiated the study. L.W.~and W.X.~wrote the paper and revised it with K.S.K.~and T.-W.L. The authors declare no competing financial interests. Supplementary Information is available in the online version of the paper. Correspondence and requests for materials should be addressed to K.S.K. (kimks@unist.ac.kr) or T.-W.L. (twlees@snu.ac.kr).   
\end{acknowledgments}

\bibliography{suprabinary}

\end{document}